\input{aipcheck}

\documentclass[
    ,final            % use final for the camera ready runs
  ]
  {aipproc}

\usepackage{aas_macros}

\layoutstyle{6x9}

\begin{document}

\title{Astrophysics of white dwarf binaries}


\author{Gijs Nelemans}{
  address={Department of Astrophysics, IMAPP, Radboud University
  Nijmegen}
}

\begin{abstract}
White dwarf binaries are the most common compact binaries in the
Universe and are especially important for low-frequency gravitational
wave detectors such as LISA. There are a number of open questions
about binary evolution and the Galactic population of white dwarf
binaries that can be solved using gravitational wave data and at the
same time, our ever improving knowledge about these binaries will help
to predict the signals that can be expected for LISA. In addition a
number of white dwarf binaries will serve as verification sources for
the instrument. I will discuss these issues and report recent,
surprising, developments in this field. Finally I report calculations
about the feasibility of complementary electro-magnetic observations
which unfortunately cannot reproduce the optimistic results of
\citet{cfs03}.
\end{abstract}

\maketitle

%%%%%%%%%%%%%%%%%%%%%%%%%%%%%%%%%%%%%%%%%%%%
%% MAINMATTER
%%%%%%%%%%%%%%%%%%%%%%%%%%%%%%%%%%%%%%%%%%%%

\section{Introduction: white dwarf binaries}

White dwarf binaries are the most common compact binaries in the
Universe, simply because the vast majority of stars that evolve to
become a compact object form a white dwarf. In addition most stars (as
far as we know) are formed in binary systems, roughly half of which
have orbital periods short enough that the evolution of the two stars
when they become giants is strongly influenced by the presence of a
companion, either through indirect influence of tides -- causing
changes in rotation and possibly enhancing the stellar wind -- or
directly through mass exchange.

In particular it has become clear from observed close binaries, that a
large fraction of binaries that interacted in the past must have lost
considerable amounts of angular momentum, thus forming compact
binaries, with compact stellar components \citep[e.g.][]{pac76}. The
details of the evolution leading to this loss of angular momentum are
uncertain, but generally this is interpreted in the framework of the
so called ``common-envelope evolution'': the picture that in a
mass-transfer phase between a giant and a more compact companion the
companion quickly ends up inside the giant's envelope, after which
frictional processes slow down the companion and the core of the
giant, causing the ``common envelope'' to be expelled, as well as
the orbital separation to shrink dramatically \citep[e.g.][]{ts00b}.

If we restrict ourselves to the most compact binaries know, often
called ultra-compact binaries, in which both components of the binary
are compact objects, we distinguish two types of binaries:
\emph{detached} binaries, in which the two components are relatively
widely separated and \emph{interacting} binaries, in which mass is
transferred from one component to the other. In the detached class, we
will be concentrating on double white dwarfs, while in the interacting
class, there are two types: white dwarfs accreting from a white
dwarf like object (the so called AM CVn systems, after the prototype
of the class, the variable star AM CVn) and neutron stars accreting
from a white dwarf like object, known as ultra-compact X-ray binaries
(UCXBs) \citep[e.g.][]{war95,nel05,vh95}.

\section{Astrophysics with double white dwarfs}

Ultra-compact binaries are interesting objects for a number of
reasons. A very brief sketch of some of the most important
astrophysical questions related to double white dwarfs follows.
\begin{description}
\item[Binary evolution] Double white dwarfs are excellent tests of
  binary evolution. In particular the orbital shrinkage during the
  common-envelope phase can be tested using double white dwarfs. The
  reason is that for giants there is a direct relation between
  the mass of the core (which becomes a white dwarf and so its mass is
  still measurable today) and the radius of the giant. The latter
  carries information about the (minimal) separation between the two
  components in the binary before the common envelope, while the
  separation after the common envelope can be estimated from the
  current orbital period. This enables a detailed reconstruction of
  the evolution leading from a binary consisting of two main sequence
  stars to a close double white dwarf \citep{nvy+00}. The interesting
  conclusion of this exercise is that the standard schematic
  description of the common envelope \citep{pac76} -- in which the
  envelope is expelled at the expense of the orbital energy -- cannot
  be correct. An alternative scheme, based on the angular momentum,
  for the moment seems to be able to explain all the observations
  \citep{nt03}.

\begin{figure}
  \includegraphics[height=7.5cm,angle=0]{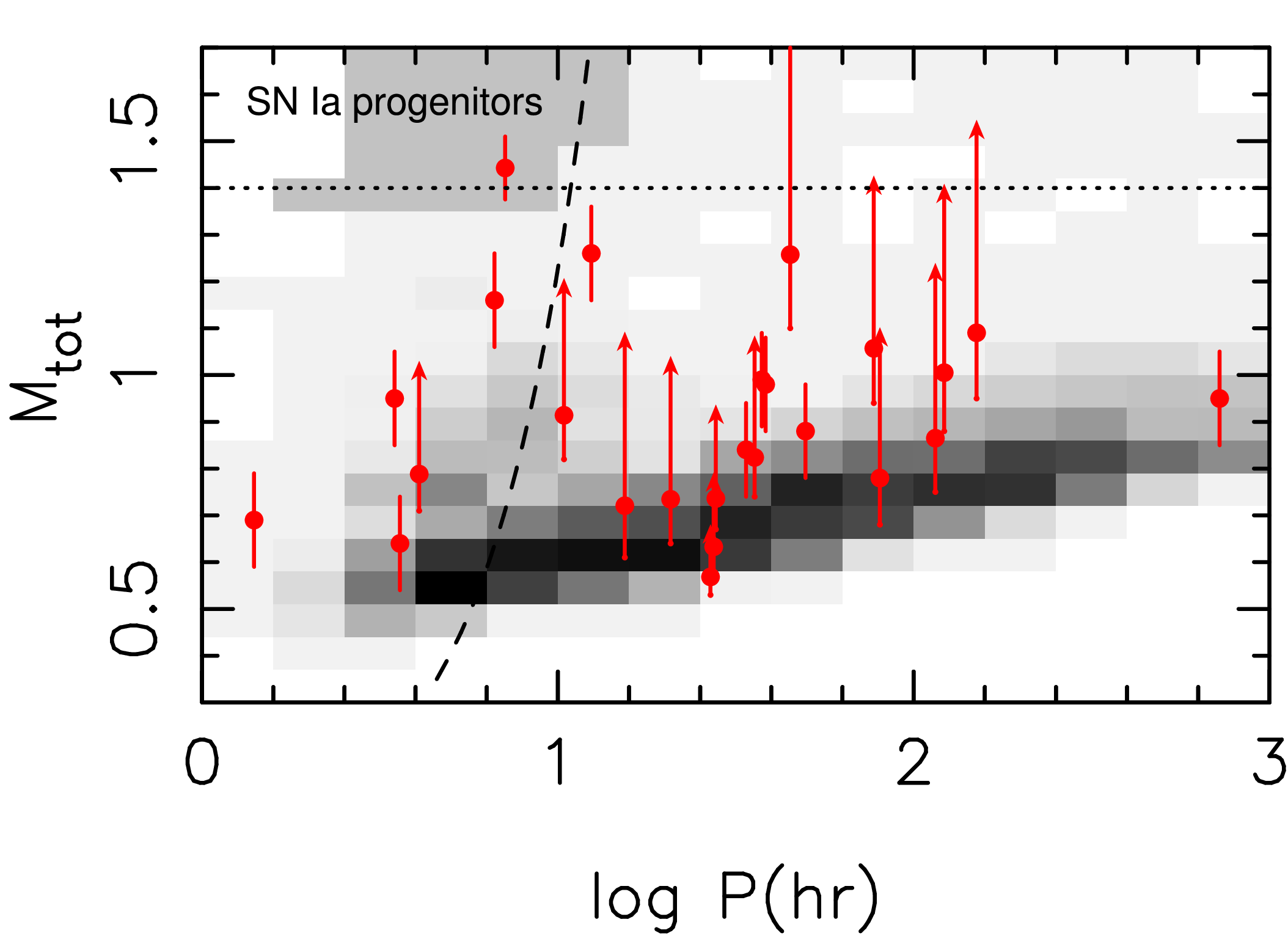}
  \caption{Period versus total mass of double white dwarfs. The points
  and arrows are observed systems \citep{nnk+05}, the grey shade a
  model for the Galactic population. Systems to the left of the dashed
  line will merge within a Hubble time, systems above the dotted line
  have a combined mass above the Chandrasekhar mass. The top left
  corner shows the region of possible type Ia supernova progenitors,
  where the grey shade has been darkened for better visibility.  }
\label{fig:P_Mtot}
\end{figure}

\item[Type Ia supernovae] Type Ia supernovae have peak
  brightnesses that are well correlated with the shape of their light
  curve \citep{phi93}, making them ideal standard candles to determine
  distances. The measurement of the apparent brightness of far away
  supernovae as a function of redshift has led to the conclusion that
  the expansion of the universe is accelerating
  \citep{pad+98,rst+04}. This depends on the assumption that these
  far-away (and thus old) supernovae behave the same as their local
  cousins, which is a quite reasonable assumption. However, one of the
  problems is that we do not know what exactly explodes and why, so
  the likelihood of this assumption is difficult to assess
  \citep[e.g.][]{pml+06}. One of the proposed models for the
  progenitors of type Ia supernovae are massive close double white
  dwarfs that will explode when the two stars merge \citep{it84a}.
  The SPY survey \citep{ncd+01} is aimed at making an inventory of
  double white dwarfs to test this proposal, at least in terms of
  whether the expected number of mergers is compatible with the type
  Ia supernova rate. In Fig.~\ref{fig:P_Mtot} the observed double
  white dwarfs are compared to a model for the Galactic population of
  double white dwarfs \citep{nyp+00}, in which the merger rate of
  massive double white dwarfs is similar to the type Ia supernova
  rate. The grey shade in the relevant corner of the diagram is
  enhanced for visibility. The discovery of at least one system in
  this box confirms the viability of this model (in terms of event
  rates). See \citet{tou05} for a review of type Ia progenitor models.
\item[Accretion physics] The fact that in AM CVn systems, as well as
  UCXBs the mass losing star is an evolved, hydrogen deficient star,
  gives rise to a unique astrophysical laboratory, in which accretion
  discs made of almost pure helium, or in the case of UCXBs, of almost
  pure carbon and oxygen are formed
  \citep[e.g.][]{mhr91,sch+01,gns+01,rgm+05c,wnr+06,njs06}. This opens
  the possibility to test the behaviour of accretion discs of
  different chemical composition.
\end{description}

\section{Relevance for/of LISA}

The compact binaries described above have a number important
connections with LISA.
\begin{enumerate}
\item Firstly, there are a number of binaries known that should be
  detected by LISA, probably within a few weeks/months, the
  \emph{verification binaries} \citep{phi02,sv06}. The LISA
  measurements will provide additional information to the already
  known parameters of these systems.
\item Current estimates of the Galactic population of these binaries
  predict tens of millions in the LISA frequency band, so many that
  they will form an unresolved background signal that can form an
  additional noise component in the instrument
  \citep[e.g.][]{eis87,hbw90}.
\item LISA is expected to individually detect thousands of compact
  binaries throughout the Galaxy \citep[e.g.][]{eis87,nyp01}, which
  will allow a completely new and complementary way to study Galactic
  populations of compact binaries and Galactic structure.
\end{enumerate}

However, our knowledge on all these topics is fairly limited. This had
led to a number of initiatives to try to come to a better
understanding of the Galactic population of compact binaries in the
Galaxy and the science that can be done with gravitational wave
measurements, well before the launch of LISA in order to develop the
best strategies for LISA data analysis and complementary
electro-magnetic observations.

\section{Current developments}

Current developments include both theoretical investigations, new
observations as well as studies to investigate the interplay between
gravitational wave measurements and more traditional electro-magnetic
observations. I will discuss a number of these issues.

\subsection{Verification binaries}

\begin{figure}
  \includegraphics[height=12cm,angle=-90]{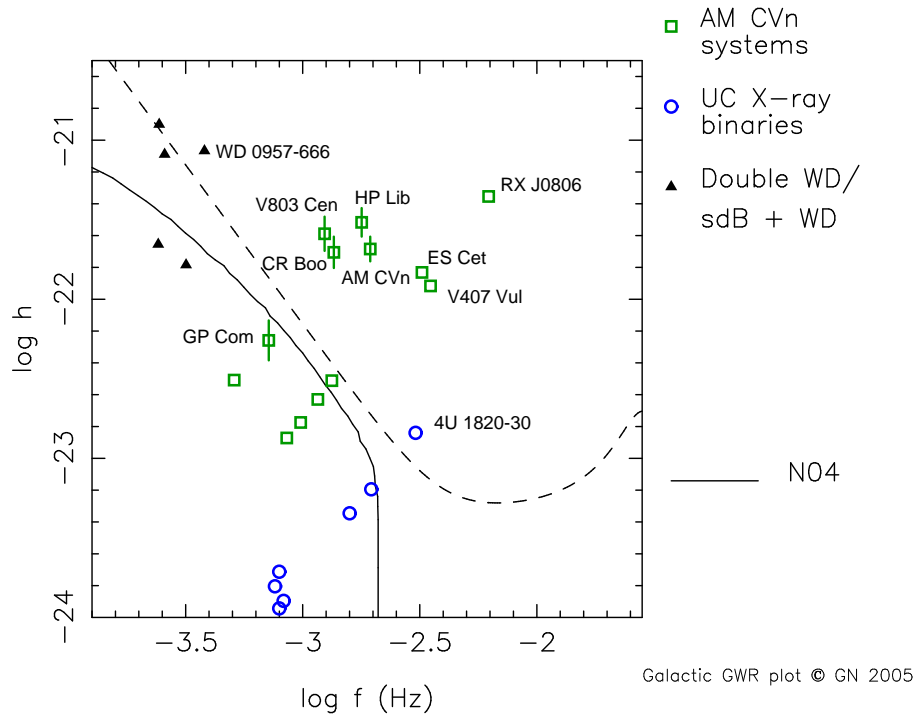}
  \caption{Expected signals of ultra-compact binaries,
  the ones with error bars from \citet{rgb+06}}
\label{fig:fh_HST}
\end{figure}

In the last year, the prototype of the interacting double white
dwarfs, the variable star AM CVn, has been studied in detail using the
William Herschel Telescope and the Hubble Space Telescope
\citep{rgn+06,rgb+06}. AM CVn is one of the Verification sources
\citep{phi02,sv06}. The new data yielded two surprises, the first is
that the distance to the system is much larger than was expected: 606
pc, rather than the estimated 235 pc \citep{rgb+06}. The second was
the fact the the mass of the donor star and the mass ratio of the
system are larger than expected: rather than fully degenerate, the
donor turns out to be semi-degenerate \citep{rgn+06}. Together these
changes leave the expected signal strength as will be measure with
LISA rather unchanged, but this expectation is now based on firm
measurements and a realistic error bar can be given (see
Fig.~\ref{fig:fh_HST}). Similar arguments for other systems (HP Lib,
CR Boo, V803 Cen and GP Com) give results also plotted in
Fig.~\ref{fig:fh_HST}, yielding a total of four definite verification
binaries. The two shortest period systems, RX~J0806.3+1527 and V407
Vul are still much debated and their periods are not yet confirmed as
orbital and their masses are very uncertain
\citep[e.g.][]{str05,icd+04,rhw+05,mn05,dis06,dt06}

\subsection{Galactic white dwarf background}

\begin{figure}
  \includegraphics[height=7cm,angle=0]{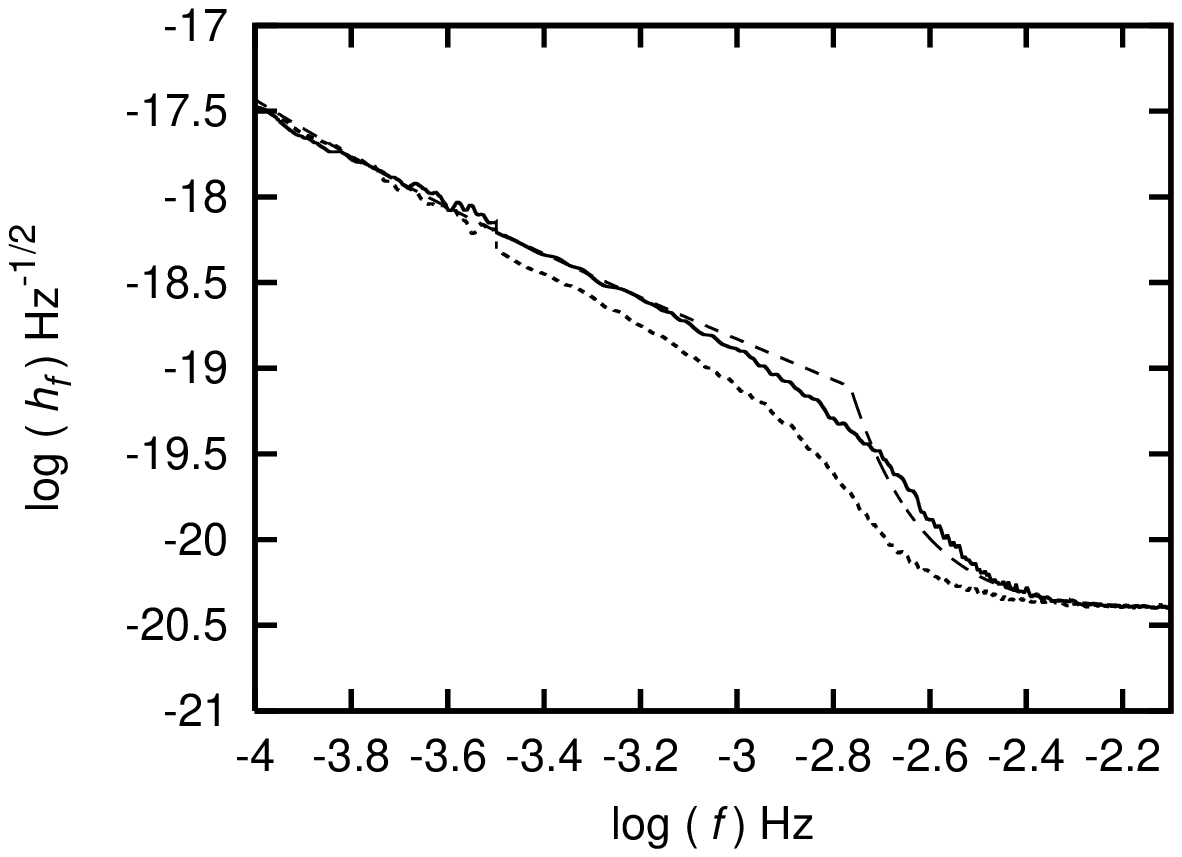}
  \caption{Comparison of the backgrounds based on the population
  models of \citet{nyp03} (solid line) and 10\% \citet{hbw90} (dotted
  line) and the \citet{bc04} estimate. From \citet{trc05}} 
\label{fig:bg}
\end{figure}

Another, rather amusing, ``new result'' concerns the Galactic
gravitational wave background. The original \citet{hbw90} article
(HBW90) presents the double white dwarf background in a rather generic
way, which made it difficult to compare with more recent population
synthesis based backgrounds such as my 2001 model \citep{nyp01}. The
worrying aspect was always that the HWB90 background was comparable to
my 2001 background, but seemingly contained much less sources. The
recent detailed recalculation of the HBW90 population \citep{trc05}
allowed a direct comparison. This comparison, made last year at the
Aspen conference, yielded the surprising result that the fact that
indeed there are about a factor of 10 less sources (resulting in a
$\sqrt{10}$ lower background) was almost exactly offset by the larger
chirp masses in the HBW90 model compared to my 2001 model \citep[see
also][and Fig.~\ref{fig:bg}]{trc05}. The first is due to the, at that
time, almost complete lack of observed close white dwarf binaries that
led to the famous factor 10 reduction in the number of binaries
advocated by HBW90. The higher chirp mass is due to the fact that
HBW90 consider only one of the formation channels to the formation of
close double white dwarfs \citep[see][for a description of the
formation channels]{nyp+00} and modelled this in a simple way which
led to binaries typically consisting of a low-mass white dwarf with a
massive ($\sim 1 M_\odot$) companion. The later models, and the
currently known double white dwarfs, predominantly contain binaries
with two low-mass white dwarfs \citep[e.g.][]{mar00,nyp+00,nnk+05}. As
the observational data now clearly confirm the dominance of low-mass
systems, the reduction of the number of systems in HBW90 thus is a
lucky ``mistake'', yielding the right results for the wrong reason!

\subsection{Complementary electro-magnetic observations}

\begin{figure}
  \includegraphics[height=7cm,angle=0]{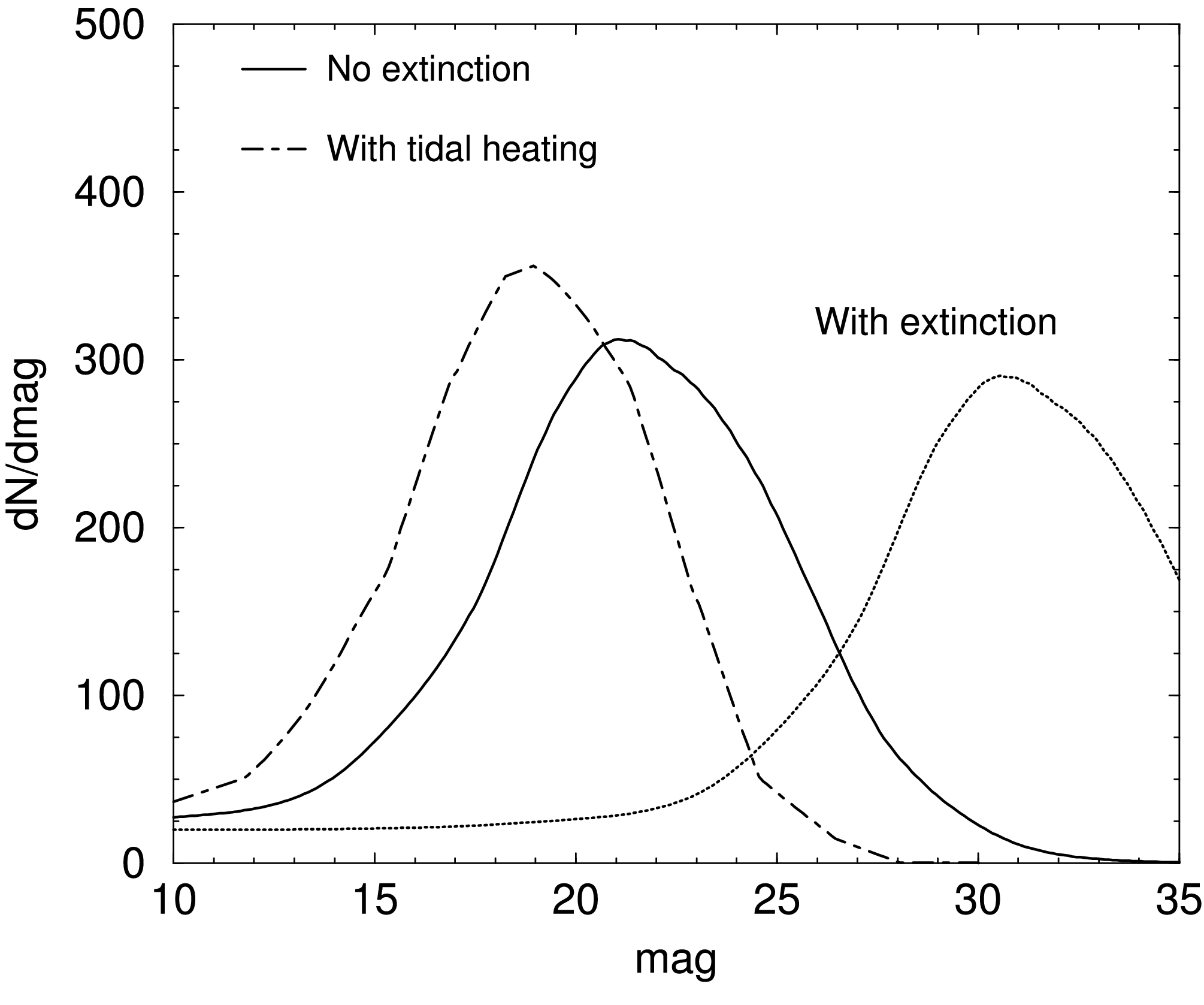}
  \caption{Distribution of expected V-band magnitudes of resolved LISA
  sources according to \citet{cfs03}} 
\label{fig:cooray}
\end{figure}

\begin{figure}
  \includegraphics[height=11cm,angle=-90]{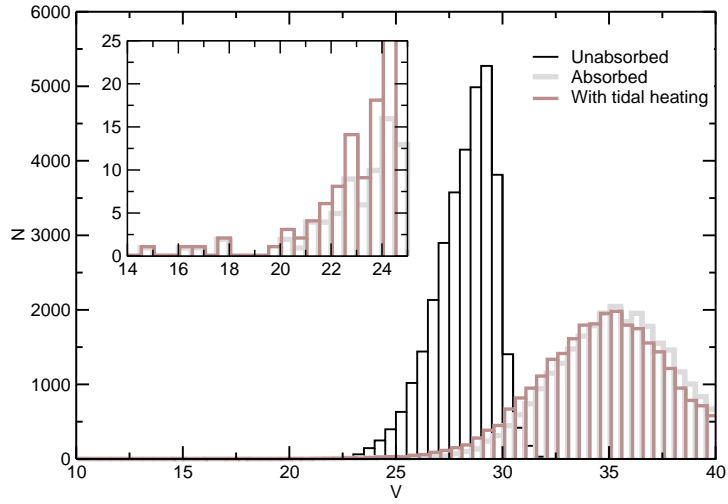}
  \caption{Histogram of expected V-band magnitudes of resolved LISA
  sources according to the model described in \citet{nyp03}. Tidal
  heating according to Fig. 6 of \citep{itf98}.}
\label{fig:dwd_hist}
\end{figure}

\begin{figure}
  \includegraphics[height=12cm,angle=-90]{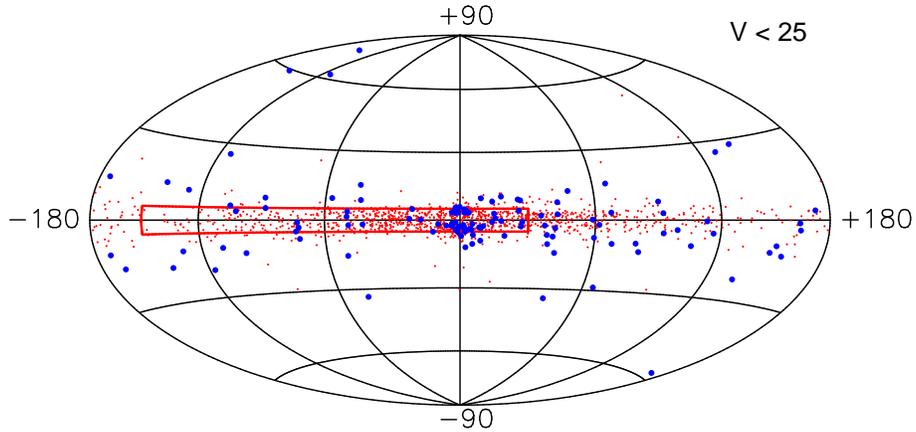}
  \caption{Sky position of the expected resolved LISA systems that can
  be observed with V $<$ 25, in Galactic coordinates. The detached
  double white dwarfs are plotted as the large symbols, the AM CVn
  systems as the small dots. The area that will be covered by
  EGAPS-south is indicated.}
\label{fig:lb}
\end{figure}

My final section is devoted to an issue which I think at the moment
deserves much more attention than it receives: the investigation of
the importance of complementary electro-magnetic
observations. \citet{cfs03} studied the feasibility of observing the
resolved LISA sources with optical telescopes and concluded that ``up
to many tens of percent'' of the resolved sources might be detectable
with V-band magnitudes between 20 and 25 (see
Fig.~\ref{fig:cooray}). This is a very promising result, but
unfortunately I think it cannot be true. The simple reason being that
typical white dwarfs have absolute magnitudes above 10 and the vast
majority of the resolved systems will be close to the Galactic centre
with a distance modulus of about 15, so the majority of the systems
should have magnitudes above 25 even before absorption is taken into
account. A histogram of the expected magnitudes for the resolved
double white dwarf from my 2004 calculation \citep{nyp03} is shown in
Fig.~\ref{fig:dwd_hist} and shows that after taking absorption into
account only a tiny fraction of resolved LISA sources is expected to
have a V-band magnitude below 25.  \citet{itf98} looked at the
possible heating of double white dwarfs when their orbital periods
become very short. The resolved LISA sources fall exactly in this
class. Although the maximum tidal heating luminosity can be quite
substantial, the increase in surface luminosity for the typically
low-mass close white dwarf pairs is rather modest (see Fig.~6 of
\citet{itf98}), so tidal heating is likely not going to change this
picture dramatically (see Fig.~\ref{fig:dwd_hist}). 

The prospects of detecting the resolved interacting AM CVn systems
with optical and/or X-ray instruments are more promising
\citep{nyp03}, due to the fact that the sort period AM CVn systems
have high mass-transfer rates and thus are bright. In
Fig.~\ref{fig:lb} I show the expected number of systems detectable
(taken to be V $<$ 25) plotted on the sky in Galactic coordinates. The
systems show a clear concentration in the Galactic plane. The area
that will be observed using the ESO VST telescope as part of the
European Galactic Plane Surveys (EGAPS)
\footnote{http://www.egaps.org} is indicated. A total of about 750 AM
CVn systems and 50 double white dwarfs are expected.

\section{Conclusions}

Recent developments provide much improved estimates of the parameters
of some verification binaries, with AM CVn, HP Lib, CR Boo and V803
Cen now all being definite verification binaries. A better
understanding of the \citet{hbw90} calculation shows that their rather
arbitrary reduction of the number of expected sources by a factor of
10 was a lucky move to offset the overestimate of the typical chirp
mass of a double white dwarf by roughly a factor 2. Finally, the
optimistic conclusion by \citet{cfs03} that maybe tens of percent of
the resolved LISA double white dwarfs can be observed optically seems
literally too good to be true and unless I'm very much mistaken only
several tens of resolved double white dwarfs will have V $<$
25. Several hundred resolved AM CVn systems are expected to be
detectable optically. However, all these calculations are still very
simple and determining reliable estimates of what can and should be
done in terms of electro-magnetic preparation and follow-up of
(resolved) LISA sources is one of the top priorities for the next
years.

%%%%%%%%%%%%%%%%%%%%%%%%%%%%%%%%%%%%%%%%%%%%%%%%
%% BACKMATTER
%%%%%%%%%%%%%%%%%%%%%%%%%%%%%%%%%%%%%%%%%%%%%%%%

\begin{theacknowledgments}
 It is a pleasure to thank all my colleagues for stimulating
 discussions. This work is supported by NWO-VENI grant 639.041.405.
\end{theacknowledgments}



\bibliographystyle{aipproc}   % if natbib is available

\bibliography{journals,binaries}

\IfFileExists{\jobname.bbl}{}
 {\typeout{}
  \typeout{******************************************}
  \typeout{** Please run "bibtex \jobname" to optain}
  \typeout{** the bibliography and then re-run LaTeX}
  \typeout{** twice to fix the references!}
  \typeout{******************************************}
  \typeout{}
 }

\end{document}